# Impact of Mobility and Transmission Range on the Performance of Backoff Algorithms for IEEE 802.11-Based Multi-hop Mobile Ad hoc Networks

Pooja Saini
Department of Computer Engineering,
SDDIET, PKL, INDIA
Email-Id : apspst.09@gmail.com

*Abstract*

The medium access control (MAC) protocol is the main element which determines the system performance in wireless local area networks. The MAC technique of the IEEE 802.11 protocol is called Distributed Coordination Function (DCF). In IEEE 802.11 Wireless Local Area Networks (WLANs), network nodes experiencing collisions on the shared channel need to backoff for a random period of time, which is uniformly selected from the Contention Window (CW). This contention window is dynamically controlled by the Backoff algorithm. First step to design a an efficient backoff algorithm for multi-hop ad hoc network is to analysis of the existing backoff algorithms in multi-hop ad hoc networks. Thus, in this paper, we considered two important multi-hop adhoc network scenarios: (a) *Node Mobility Scenario* and (b) *Transmission Range Scenario* and analyze and evaluate both the impact of mobility (i.e. node speed) and the impact of transmission range of nodes on the performance of various backoff algorithms.

## 1. Introduction

A Mobile Ad hoc Network (MANET) [1] is a dynamic wireless network that is established by a group of mobile stations without necessarily using pre-existing infrastructure or centralized administration. Such networks can be useful in disaster recovery where there is not enough time or resources to configure a wired network. The IEEE 802.11 WLAN MAC/PHY specification [2] is one of the recommended international standards for WLANs. The standard contains technical details for the Medium Access Control layer (MAC) and the Physical layer (PHY) of the communication protocol. Two coordination functions are defined in the IEEE 802.11 MAC/PHY standard: the *Point Coordination Function (PCF)* and the *Distributed Coordination Function (DCF)*. In the PCF mechanism, a polling technique is employed by the access points or base stations to query network nodes for any traffic they may have to send. In the DCF medium access mode, active nodes compete for the use of the channel in a distributed manner via the use of the Carrier Sensing Multiple Access with Collision Avoidance (CSMA/CA) scheme. Packet collisions are not completely eliminated in the IEEE 802.11 MAC/PHY standard due to the distributed nature of the competing nodes and the bursty traffic arrival at the nodes. In the IEEE 802.11 DCF scheme, the senders of the colliding packets need to refrain from immediate retransmissions in order to avoid repeated collisions. Thus, each competing node sets up a backoff timer according to a randomly selected backoff time period and enters the backoff state. This backoff time period is selected uniformly between 0 and the Contention Window (CW). In the IEEE 802.11 DCF scheme, the CW is dynamically controlled by the backoff algorithm; the *Binary Exponential Backoff (BEB)*. In the BEB algorithm, the contention window is doubled every time a node experiences a packet collision. If a node is successful in its packet transmission, the contention window is reset to the minimum value. In order to avoid the contention window from growing too large or shrinking too small, two bounds on CW are defined: the maximum contention window (CWmax) and the minimum contention window (CWmin). However, the BEB scheme suffers from a fairness problem; some nodes can achieve significantly larger throughput than others. The fairness problem occurs due to the fact that the scheme resets the contention window of a successful sender to CWmin, while other nodes continue to maintain larger contention windows, thus reducing their chances of seizing the channel and resulting in channel domination by the successful nodes.

Various backoff algorithms have been proposed. In [3], multiplicative increase linear decrease (MILD) algorithm is proposed where a node increases its backoff interval by 1.5 after every unsuccessful transmission and decreases its backoff interval by one after successful transmission. In [4], exponential increase exponential decrease (EIED) backoff algorithm is proposed to enhance the performance of DCF. In this scheme, a node increases its backoff interval by $r_I$ (typical values of $r_I$ are 2, $2\sqrt{2}$) after every unsuccessful transmission and decreases its backoff interval by $r_D$ (typical values of $r_D$ are $2^{1/2}$, $2^{1/4}$, 2, $2\sqrt{2}$).

In [5], modified BEB algorithm has been proposed. In this, the backoff time is increased exponentially, but with a reduced base value (less than 2) after each unsuccessful transmission until prescribed maximum value (CWmax) is reached. Whenever a node transmits a packet successfully, backoff time is reduced to a specified minimum value (CWmin). In [6], logarithmic backoff algorithm has been proposed that uses logarithmic increment of window size.

In [7], double increment and double decrement backoff algorithm has been proposed. In this algorithm, a node increases its backoff interval by 2 after each unsuccessful transmission and decreases its backoff interval by half after successful transmission.

The analysis of the existing backoff algorithms in multi-hop ad hoc networks is the first step to designing an efficient backoff algorithm for multi-hop ad hoc network. Thus, in this paper, we considered two important multi-hop adhoc network scenarios: (a) *Node Mobility Scenario* and (b) *Transmission Range Scenario* and analyze and evaluate both the impact of mobility (i.e. node speed) and the impact of transmission range of nodes on the performance of six backoff algorithms namely, BEB, Modified BEB, MILD, EIED, DIDD and Logarithmic. We have chosen only these six backoff algorithms as they belong to same category, operation-wise.

Rest of the paper is organized as follows. In Section II, we describe simulation methodology. In Section III the performance of backoff algorithms is evaluated and compared. We finally draw our conclusions in Section IV.

## 2. Simulation Methodology

Simulation studies have been carried out using GloMoSim [8] network simulator which allows node mobility, thereby providing simulation of MANETs. Our simulation considered a network of 50 mobile nodes placed randomly within a 1000 x 1000 m$^2$ area. Constant bit rate (CBR) data sessions among randomly chosen source-destination pairs (SDPs) are used. For example, with 10 SDPs amongst 50 nodes, 10 source nodes and 10 destination nodes (i.e., 20 nodes in total) will be engaged in data transfer. However, during this data transfer process, all of the 50 nodes (including the above 20 nodes) will operate in the background for providing necessary support (i.e., routing/forwarding) to the ongoing communication process in the network.

*(a) Node Mobility Scenario*: Node movement is modeled using the random waypoint mobility model (RWMM), which is widely used in MANET simulations. In RWMM, nodes move at a speed uniformly distributed in [MIN SPEED, MAX SPEED]. Each node begins the simulation by moving towards a randomly chosen destination. Whenever a node reaches a destination, it rests for a pause time. It then chooses a new destination and moves towards the same. This process is repeated until the end of simulation time. In our simulations, We considered 5m/s, 10m/s, 15m/s, 20m/s, 25m/s and 30m/s as average node speed and also the pause time is set at zero (i.e., nodes move continuously throughout the simulation period). This is done to study the impact of continuous node mobility (i.e., worst-case scenario) on the network performance.

(b) *Transmission Range Scenario*: Transmission range of a node refers to the average maximum distance in usual operating conditions between two nodes. We can change the radio range by varying the transmitter power (RADIO-TX-POWER) or the receiver power (RADIO-RX-THRESHOLD), it is somehow advisable to change the transmitter power, because the receiver power depends of the radio environment while we can control the transmitter power. We considered 50, 100, 150, 200, 250 and 300 meters as radio ranges.

The data rate is 2 Mbps while the data packet size is 512 bytes. The data packets are sent at a rate of 4 packets /sec by each source. Each simulation is executed for 30 minutes. However, data packets are generated by CBR sources only during last 800 seconds of simulation time. To avoid initial transient problem and the problem with RWMM model as reported in [9], in our simulations we discard the initial 1000 seconds of simulation period. Five runs with different seeds have been conducted for each scenario and collected data is averaged over these runs. A summary of multi-hop network salient simulation parameters for both scenario are presented in Table I.

## 3. Results and Discussion

Performance of the algorithms is evaluated in terms of packet delivery ratio, average end-to-end packet delay as defined in the following:

*Packet delivery ratio (PDR)*: The ratio of the data packets delivered to the destinations to those generated by the CBR sources. Mathematically, we can define as,

$$\text{PDR} = \frac{\left[\sum_{i=1}^{m} \frac{Sum\ of\ data\ packets\ received\ by\ each\ destination}{Sum\ of\ data\ packets\ generated\ by\ each\ CBR\ source}\right]}{m}$$

where,

  i, indicates the current output file
  m, indicates the total number of output files

Table 1 : Salient Simulation Parameters

| Parameter | Value |
|---|---|
| Simulation Time | 30 minutes |
| Terrain Area | $1000 \times 1000$ m$^2$ |
| No. of nodes | 50 |
| Node placement strategy | Random |
| Mobility model | Random Waypoint |
| Speed of a mobile | Uniformly distributed in [MIN SPEED, MAX SPEED] |
| Pause time | 0 second |
| Propagation model | Free Space |
| Channel frequency | 2.4 GHz |
| Data rate | 2 Mbps |
| Radio type | Accumulated noise model |
| Network protocol | IP |
| No. of SDPs | 25 |
| MAC protocol | IEEE 802.11 DCF with CWmin =32, CWmax = 1024 |
| Routing protocol | Ad hoc on-demand distance vector (AODV) |

*Average end-to-end delay of data packets*: The average delay a data packet takes to travel from the source to the destination node. This includes all possible delays caused by buffering during route discovery latency, queuing at the interface queue, retransmission delays at the MAC, and propagation delay. Mathematically, we can define as,

$$\text{Average end–to-end delay} = \frac{\left[\sum_{i=1}^{m} Sum\ of\ average\ end-to-end\ delay\ for\ each\ destination\right]}{m}$$

We have simulated six backoff algorithms (BEB, Modified BEB, MILD, EIED, DIDD and logarithmic) in single-hop as well as in multi-hop ad hoc network environment with same network setting. While simulating EIED, we have chosen the best values of $r_I$ (i.e., 2) and $r_D$ (i.e., $2^{1/8}$) as indicated in [9] and for modified BEB, we have chosen the best value of b (i.e., 1.5) as indicated in [ 2, 3 & 6]. The impact of node mobility and offered load (i.e., number of SDPs) on the network performance in multi-hop ad hoc network is shown in Fig. 1 and Fig.2, in terms of packet delivery ratio and average end-to-end delay. And, the impact of transmission range of nodes and offered load on the network performance in multi-hop ad hoc network is shown in Fig. 3 and Fig. 4, in terms of packet delivery ratio and average end-to-end delay.

In Fig. 1, we observe the impact of node mobility (i.e., node speed) and offered load on packet delivery ratio (PDR). In general, it is noticed that the PDR decreases as the node speed increases. PDR is 5% to 20% better with modified BEB (b = 1.5), for higher node speeds (i.e., > 10m/s), as compared to MILD, EIED, BEB, DIDD and logarithmic. However, for low mobility (i.e., node speed ≤ 5m/s), EIED, DIDD and BEB show similar performance as modified BEB. This is because the network topology does not change significantly when the nodes are less mobile.

In Fig. 2, we observe that the modified BEB performs significantly better in terms of average end-to-end delay, for higher node speeds (i.e., > 10m/s), as compared to MILD, EIED, logarithmic, DIDD and BEB. For low mobility (i.e., node speed ≤ 5m/s), logarithmic and BEB show the similar performance as modified BEB. This is due to following reasons. The BEB, MILD, EIED, DIDD and logarithmic algorithms cause a fast growth-rate of waiting times spreading the backlog traffic over a larger time frame. However, this fast growth-rate of waiting time with increasing number of occurrence of collisions might not be appropriate for a MANET, wherein the contending nodes might leave the geographical location of contention itself after a short while due to their mobility. And also, slow-decrease in CW after a successful transmission, forces a node to go for a longer waiting time (expect in BEB and logarithmic where CW is reset to CWmin), wherein the contending nodes might move out of collision range after a short while due to their mobility.

In Fig.3, we observe the impact of transmission range of nodes and offered load i.e., 25 SDPs on packet delivery ratio. In general, it is noticed that packet delivery ratio increases with increase in transmission range of nodes. For changing the transmission range from 150m to 300m, the packet delivery ratio increases by 50%. With high transmission range (i.e., ≥ 150m), packet delivery ratio is better with modified BEB as compared to MILD, EIED, DIDD, BEB and logarithmic backoff algorithms. Figure 3 shows that, the MILD backoff algorithm performs worst and there is a difference of 15% in the performance of MILD and modified BEB. Modified BEB performs better due to slow increase in CW whenever there is an unsuccessful transmission and resetting of CW to $CW_{min}$ whenever there is a successful transmission. For low range (i.e., < 100m) the difference in packet delivery ratio is not a significant value as only few nodes are in radio range of each other.

In Fig.4, we observe the impact of transmission range of nodes and offered load i.e., 25 SDPs, on average end-to-end delay. As the transmission range of nodes increases up to 200m the average end-to-end delay also increases and after that transmission range (i.e., >200m) the average end-to-end delay starts decreasing. The average end-to-end delay decreases by 63% as the transmission range of nodes vary from 200m to 300m. For high transmission range (i.e., ≥ 150m), modified BEB performs better as compared to MILD, EIED, DIDD, BEB and logarithmic backoff algorithms. This is due to the reason that as the transmission range increases more and more nodes come in range of each other and hence the delay due to routing processing decreases and thus the average end-to-end delay decreases.

## 4. Conclusions

In this paper, we have compared the performance of BEB, modified BEB, logarithmic, EIED, DIDD and MILD backoff algorithms for IEEE 802.11 DCF based MAC protocol in multi-hop ad hoc network environment. Performance of algorithms is evaluated using simulations. Table II and Table III summarized the best and worst performance of BEB, modified BEB, MILD, DIDD, EIED and logarithmic backoff algorithms in node mobility scenario and transmission range scenario for multi-hop ad hoc network, respectively, in terms of packet delivery ratio and average end-to-end delay. We intend to explore further in this direction in our future work.

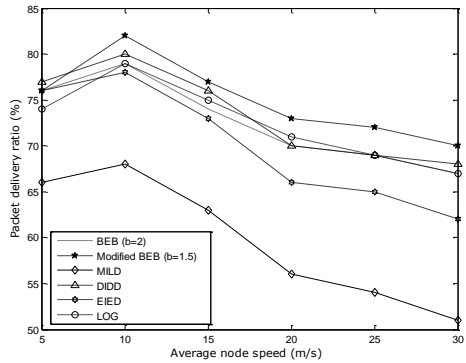

**Fig. 1. Impact of average node speed on packet delivery ratio (with 25 SDPs)**

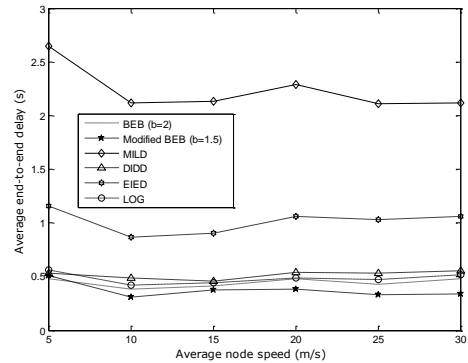

**Fig. 2. Impact of average node speed on average end-to-end delay (with 25 SDPs)**

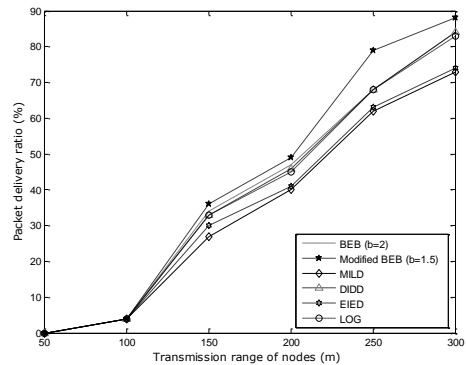

**Fig. 3 Impact of transmission range of node on packet delivery ratio (With offered load = 25 SDPs)**

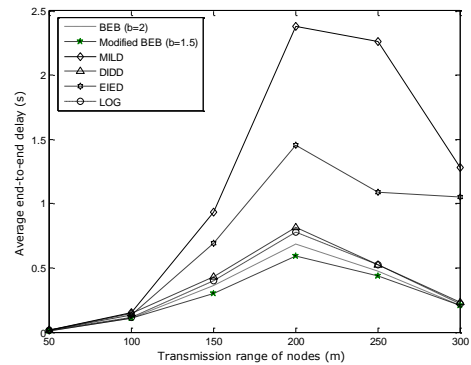

**Fig. 4 Impact of transmission range of node on average end-to-end delay (With offered load = 25 SDPs)**

Table II. Performance Analysis: Multi-hop Ad hoc Network
(With varying node mobility)

| Performance Metrics | Backoff Algorithm | | | | | | | | | | | |
|---|---|---|---|---|---|---|---|---|---|---|---|---|
| | BEB | | Modified BEB | | MILD | | DIDD | | EIED | | LOG | |
| | LM | HM | LM | HM | LM | HM | LM | HM | LM | HM | LM | HM |
| Packet delivery ratio | | | | B | | W | | | | | B | |
| Average end-to-end delay | | B | B | B | W | W | | | | | | |

where,
    B  -  Best Performance      W  -  Worst Performance
    LM  -  Low Mobility (i.e., 5 m/s)      HM  -  High Mobility(i.e., 30m/s)

International Journal of Advancements in Technology (IJoAT)  http://ijict.org/     ISSN 0976-4860

TABLE III. . Performance Analysis: Multi-hop Ad hoc Network
(With varying transmission range of nodes)

| Performance Metrics | Backoff Algorithm | | | | | |
|---|---|---|---|---|---|---|
| | BEB | Modified BEB | MILD | DIDD | EIED | LOG |
| Packet delivery ratio | | B | W | | | |
| Average end to end delay | | B | W | | | |

**Note:**
• Empty cells indicate that the performance of respective backoff algorithm is average (i.e., in between the best and worst performance).
• For some performance metrics, more than one backoff algorithm shows B ( or W ) as their respective simulation values are very close to each other.

### Acknowledgements

The author express her sincere thanks to the Director, National Institute of Technical Teachers' Training and Research, Chandigarh, India, for his support. The author also like to thank Er. Roshan Lal Jindal, Chairman, SDDIET, Panchkula, India, for his encouragement.